# Possible dwarf nova on the Digitized Sky Survey plates

ROMANOV, FILIPP DMITRIEVICH [1]

1) Moscow, Russian Federation, filipp.romanov.27.04.1997@gmail.com

**Abstract:** I report the discovery of a new bright variable star in the constellation Scorpius. It is most likely star OGLE BUL-SC23 396060 with position (J2000.0): RA 17:57:59.75, DEC -31:38:07.4. Magnitude range is 14.4 (*B*) - 21.7 (*V*). It is visible in the possible outburst on two digitized Palomar Sky Survey photographic plates, which were made 60 years ago. This object was registered as a variable star in the AAVSO VSX under the name of Romanov V1 and with UG: variability type.

## 1. Introduction

In February 2017 the author discovered a bright unknown object (in the constellation Scorpius) on the Digitized Sky Survey (DSS) plates during the search for new variable stars and using the online service DSS Plate Finder (https://archive.stsci.edu/cgi-bin/dss_plate_finder ). The object is visible in two images for the date of April 18, 1958. This is a blue POSS-I O plate (exposure time 7 minutes) and POSS-E Red Plate (exposure time 40 minutes). The object looks like a star. There was no information about variable stars in radius 50" around coordinates of this object. The MPChecker service (https://minorplanetcenter.net/cgi-bin/checkmp.cgi) shows the absence of known asteroids at the specified date in radius 5' around these coordinates. This object was registered on February 21, 2017 as a variable star in the AAVSO VSX (Watson et al. 2006) under the name of Romanov V1 (a link to the page in The International Variable Star Index: https://aavso.org/vsx/index.php?view=detail.top&oid=477039). In addition, an AAVSO Unique Identifier (AUID) of 000-BMF-566 has been assigned to the star. Fig. 1 and Fig. 2 show the images of the vicinity of this variable star.

## 2. Identification

I searched for information about this star in astronomical catalogues. I defined the position of this variable star (epoch J2000.0) from Optical Gravitational Lensing Experiment (Udalski et al. 2002) data: RA 17h57m59s.75, DEC -31h38m07s.4. The difficulty in determining the position of the star due to the fact that it has a diameter of up to 12" in the images. The most likely (by position) candidate is OGLE BUL-SC23 396060 (*V* = 21.7 mag; *I* = 19.7 mag), but it also can be OGLE BUL-SC23 392509 (*V* = 21.5 mag; *I* = 19.5 mag). Color indexes (*V* - *I* = 2.0) are the same red. The photometry of both stars in OGLE photometry database does not show any apparent light variations (see the attached light curves in Fig. 3). But it seems that star OGLE BUL-SC23 392509 shows some possible small 1-year variations of magnitude.

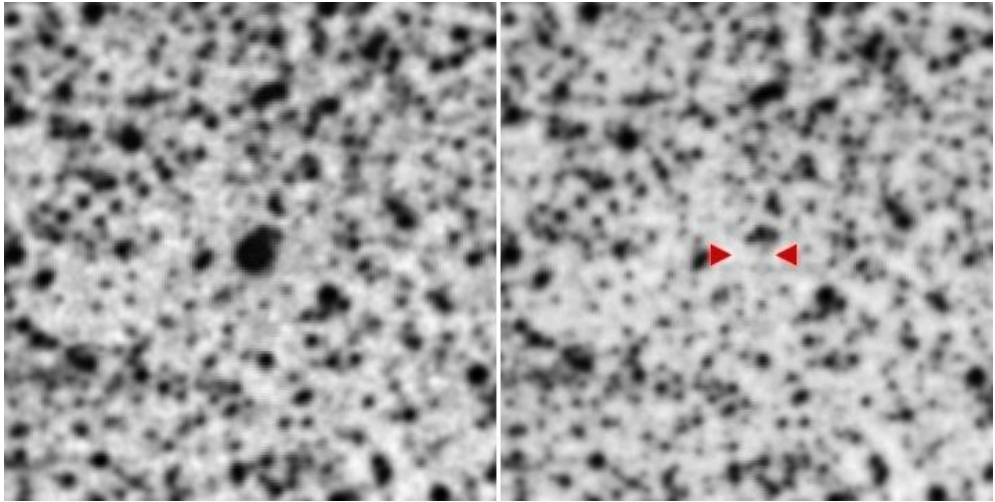

Figure 1: DSS red plates showing the vicinity of Romanov V1. Left: POSS-E Red Plate (Apr 18, 1958), right: Second Epoch Southern (UK Schmidt) plate (Jun 03, 1992). Inverted colors, field of view 3'x3'. The north is on the top, the east is on the left.

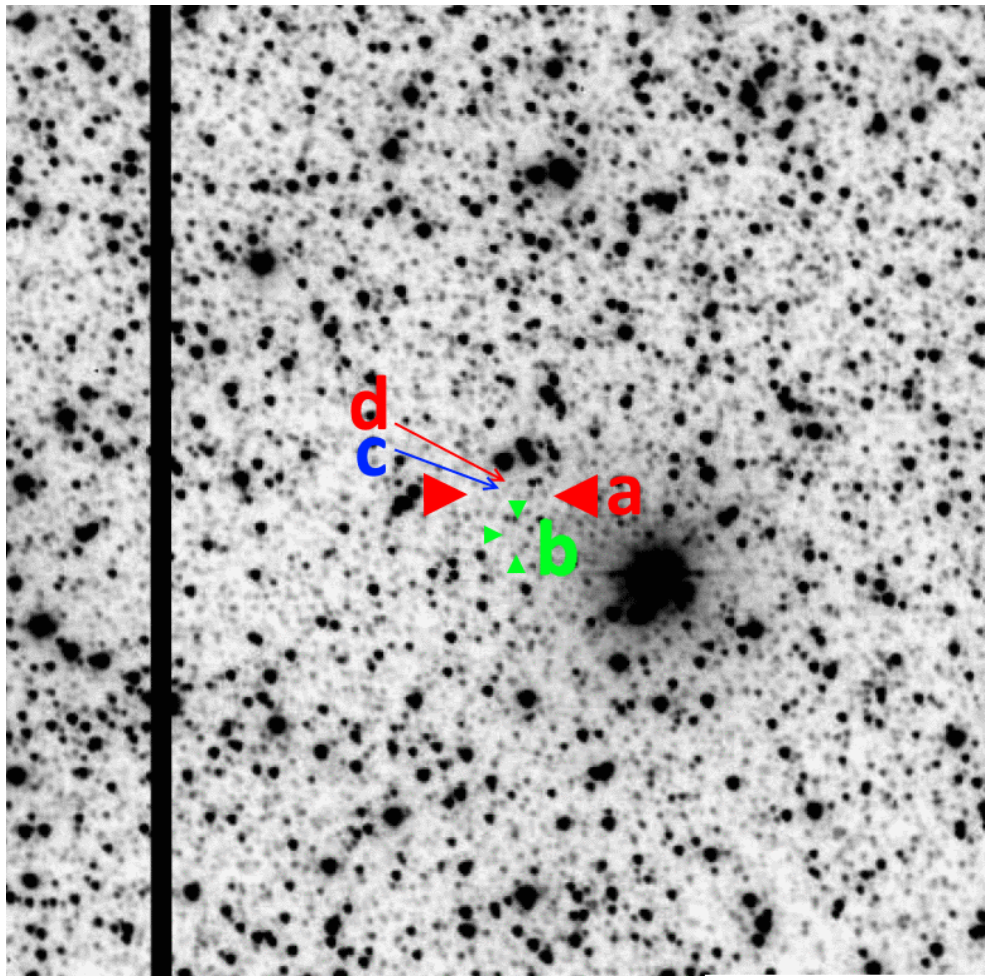

Figure 2: The frame from OGLE showing the vicinity of Romanov V1. Field of view 3'x3'. The north is on the top, the east is on the left. **a** – position of Romanov V1; **b** – position of object USNO-B1.0 0583-0618095; **c** – position of star OGLE BUL-SC23 396060; **d** – position of star OGLE BUL-SC23 392509.

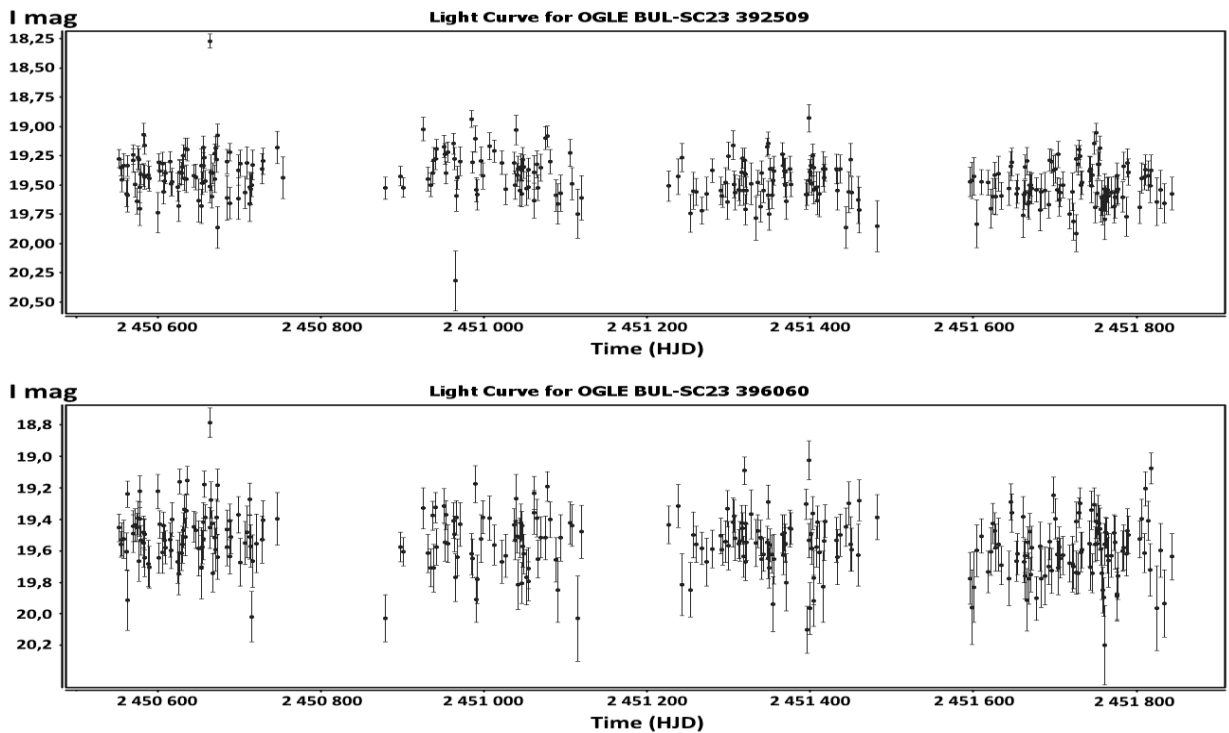

Figure 3: Light curves of OGLE BUL-SC23 392509 and OGLE BUL-SC23 396060.

There are also magnitudes in USNO-B catalog (Monet et al. 2003) for object USNO-B1.0 0583-0618095 (J2000.0 position: RA 17h57m59s.59, DEC -31h38m15s.88), which is located 9" away of this variable star. The magnitudes are 14.14 in *B* and 10.13 in *R*. The photometry was made from the same plates, on which Romanov V1 is visible. Therefore, I assume that these magnitudes were erroneously attributed to another star and refer to the variable, which I discovered.

## 3. Photometry

I determined the brightness of this variable star from two given photos in order to determine the range of its variability. In Table 1 and Table 2, there are reference stars, which I have chosen for comparison (Fig. 4 shows these stars on the blue DSS plate). *B* and *V* magnitudes (Johnson system) and *r* magnitudes (Sloan) were taken from APASS catalog: AAVSO Photometric All-Sky Survey, DR 9 (Henden et al. 2016). Also, the stars were checked for variability in the AAVSO VSX.

I did photometry with the MAXIM DL Pro Demo 6.16 software. This was hampered by the large number of background stars in the field of view of the images. As a result, I got the magnitudes of this variable star: $B = 14.4$ mag on the blue plate and $r = 10.3$ mag on the red plate. Due to the fact that the brightness of the candidate OGLE BUL-SC23 396060 is 21.7 mag in *V*, then the variability amplitude of Romanov V1 is more than 7 magnitudes, the range is 14.4 (*B*) - 21.7 (*V*).

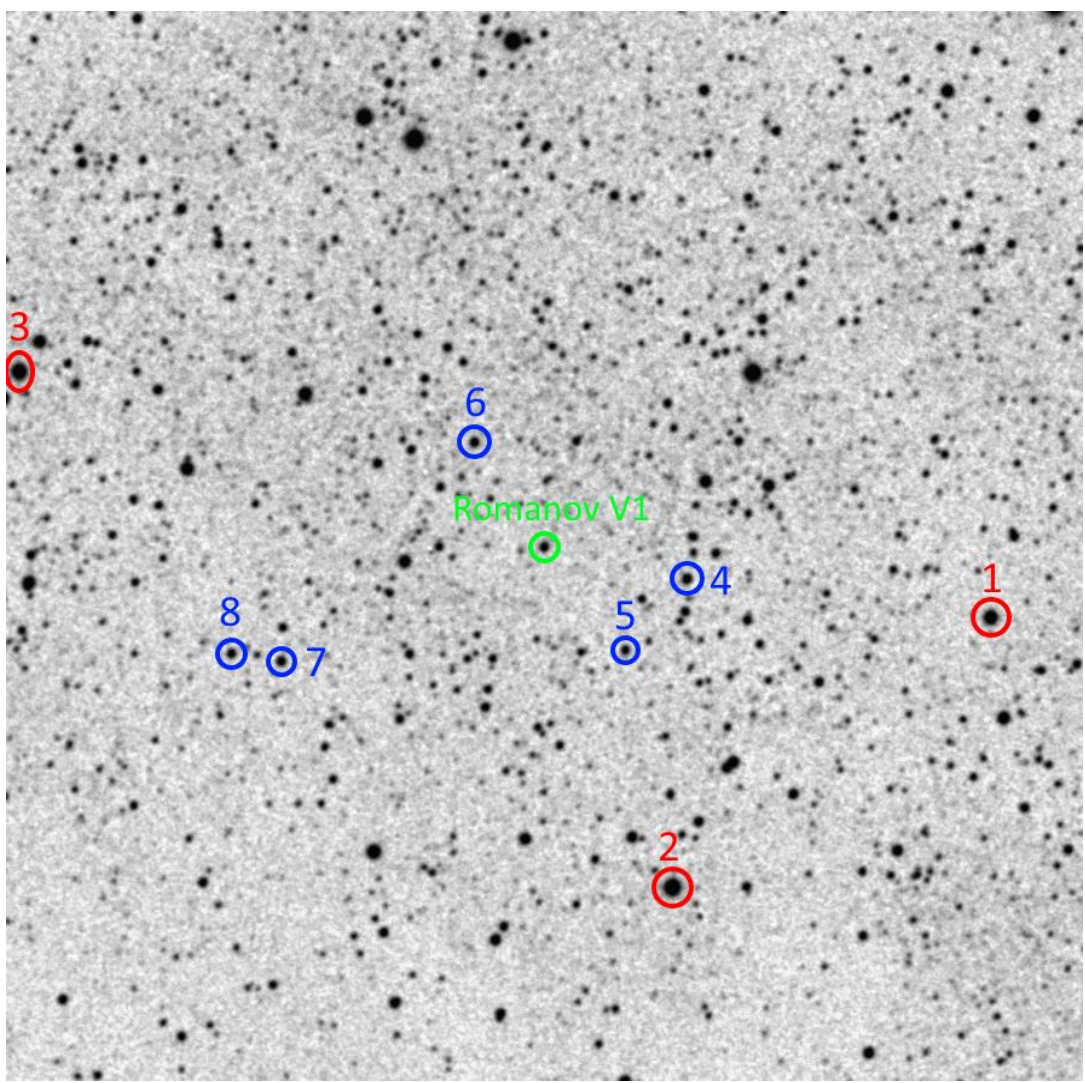

Figure 4: Romanov V1 on the blue plate POSS-I O (Apr 18, 1958). Inverted colors, field of view 15'x15'. The north is on the top, the east is on the left. Numbers denote all reference stars; the designations of these stars are in Tab. 1 and Tab. 2.

Table 1: Reference stars on the red plate.

| № from Fig. 3 | ID (UCAC4 Catalog) | RA [h m s] | DEC [° ´ ´´] | V [mag] | B [mag] | r [mag] | B - V | B - r |
|---|---|---|---|---|---|---|---|---|
| 1 | UCAC4 292-139427 | 17:57:30.621 | -31:39:12.21 | 11.561 | 11.966 | 11.499 | 0.41 | 0.47 |
| 2 | UCAC4 292-139788 | 17:57:51.678 | -31:42:57.07 | 10.237 | 10.763 | 10.113 | 0.53 | 0.65 |
| 3 | UCAC4 293-147864 | 17:58:33.943 | -31:35:34.82 | 10.816 | 11.228 | 10.751 | 0.41 | 0.48 |

Table 2: Reference stars on the blue plate.

| № from Fig. 3 | ID (UCAC4 Catalog) | RA [h m s] | DEC [° ´ ´´] | V [mag] | B [mag] | r [mag] | B - V | B - r |
|---|---|---|---|---|---|---|---|---|
| 4 | UCAC4 292-139762 | 17:57:50.454 | -31:38:36.49 | 13.505 | 14.048 | 13.313 | 0.55 | 0.74 |
| 5 | UCAC4 292-139826 | 17:57:54.556 | -31:39:35.73 | 14.319 | 14.906 | 14.012 | 0.58 | 0.89 |
| 6 | UCAC4 292-139991 | 17:58:04.248 | -31:36:39.29 | 13.828 | 14.291 | 13.723 | 0.47 | 0.57 |
| 7 | UCAC4 292-140184 | 17:58:17.066 | -31:39:41.61 | 13.609 | 14.339 | 13.387 | 0.74 | 0.95 |
| 8 | UCAC4 292-140226 | 17:58:20.328 | -31:39:34.70 | 14.013 | 14.671 | NA | 0.66 | NA |

## 4. Conclusions

From the above, known some facts about Romanov V1: variability amplitude is more than 7 magnitudes and color index is extremely red (*B-r*~4 mag). In addition, the DSS Plate Finder service has information that the blue plate was made at first (time: 10:80:00[1]) and the red plate was made later (time: 11:00:00). Exposures of these plates are 7 and 40 minutes respectively. Accordingly, it follows that this star was observed for 47 minutes (at least, because the time interval between the exposures of the plates is unknown). The star is not found on the scanned plates (approx. number of plates: 3864 from years: 1889-1891, 1893-1954, 1970-1972, 1978-1979, 1981-1989) of the Digital Access to a Sky Century at Harvard (http://dasch.rc.fas.harvard.edu ). There is no information about the outbursts of this object in the ASAS-SN database (Kochanek et al. 2017) from 2013-04-04 to the present time.

I have three hypotheses why this object has such color index. The first: it is red variable (for example, Mira variable) star. But it does not look plausible because with its position it would be a noticeable infrared source, but there is nothing apparent in the images 2MASS and WISE. Against this hypothesis, there is another objection that the star is nowhere else to be seen, except for two DSS plates.

The second: the star quickly increased its brightness, so it became brighter during the exposure of the red plate. I think that this hypothesis is incorrect, because such behavior can be observed in light curves of flare stars, but variables of this type are usually red: already written above, there are no noticeable infrared sources.

The third: the color of the star is red because of interstellar extinction (reddening). I think that this theory is the most truthful. The variable star is located only 3.7 degrees from the Galactic Center, where strong interstellar extinction takes place.

Based on the amplitude of variability and all of the above, I assume that the variable star Romanov V1 is most likely a cataclysmic variable star, namely dwarf nova of the WZ Sagittae subclass (UGWZ). However, I do not exclude that this star is a variable of another type (for example, if this star was photographed not during the maximum of its brightness). Accordingly, it is necessary to wait for the next outburst of the star to confirm or refute this. Nobody can predict the next outburst. For example, ASAS-SN is surveying the field so it can notice an outburst. I think the variable star Romanov V1 is one of the most interesting among the 45 variable stars which I discovered by now (they are registered in the AAVSO VSX). It is possible that in the archives of some observatories there are photographic plates from 1958, according to which it would be possible to perform another photometry of the star and make a light curve, according to which the type of variability will be better understood.

---

[1] It should probably be 11:20:00.

# References

Henden, A. A. et al. 2016, VizieR Online Data Catalog: II/336, 2016yCat.2336....0H

Monet, D. G. et al. 2003, *The Astronomical Journal*, 125, 984, 2003AJ....125..984M

Udalski, A. et al. 2002, *Acta Astronomica*, 52, 217, 2002AcA....52..217U

Watson, C. L. et al. 2006, 25th Annual Symp., *The Society for Astronomical Sciences*, 47, http://www.aavso.org/vsx , 2006SASS...25...47W

Zacharias, N. et al. 2012, VizieR Online Data Catalog: I/322A, 2012yCat.1322....0Z